\documentclass[aps,pre,showpacs,floatfix,twocolumn]{revtex4}
\usepackage{graphicx}
\usepackage{amsmath,amssymb}
\usepackage{natbib}
\usepackage{times}
\usepackage{hyperref}
\usepackage{color}
\usepackage{braket}

\usepackage{mathrsfs}

\begin{document}

% TITLE
\title{Protocol-Dependence and State Variables in the Force-Moment Ensemble} 
\author{Ephraim S. Bililign, Jonathan E. Kollmer, Karen E. Daniels}
\affiliation{Department of Physics, North Carolina State University, Raleigh, North Carolina 27695, USA}

\date{\today}

\begin{abstract}
Stress-based ensembles incorporating temperature-like variables have been proposed as a route to an equation of state for granular materials. 
To test the efficacy of this approach, we perform experiments on a two-dimensional photoelastic granular system under three loading conditions: uniaxial compression, biaxial compression, and simple shear. 
From the interparticle forces, we find that the distributions of the normal component of the coarse-grained force-moment tensor are exponential-tailed, while the deviatoric component is Gaussian-distributed. 
This implies that the correct stress-based statistical mechanics conserves both the force-moment tensor and the Maxwell-Cremona force-tiling area. 
As such, two variables of state arise: the tensorial angoricity ($\hat{\alpha}$) and a new temperature-like quantity associated with the force-tile area which we name {\it keramicity} ($\kappa$).
Each quantity is observed to be inversely proportional to the global confining pressure; however only $\kappa$ exhibits the protocol-independence expected of a state variable, while $\hat{\alpha}$ behaves as a variable of process.
\end{abstract}

\maketitle

% INTRODUCTION
Granular systems are characterized by the absence of thermal fluctuations and their ability to maintain a mechanically stable, jammed state in the absence of an external driving force. 
The formulation of a statistical description of jammed configurational states has been an open question for the behavior of such athermal, macroscopic particles undergoing solely repulsive contact-interactions \cite{powders, jaeger-overview, ball, martiniani}.
A promising statistical mechanics candidate, the force-moment ensemble, postulates a temperature-like state variable, conjugate to stress, called angoricity \cite{statmech}. 
For simulations of frictionless particles, angoricity has been shown to underpin an equation of state in the microcanonical ensemble \cite{henkes}, and for frictional particles, it has also satisfied a zeroth-law test \cite{puckett}.
However, the correct functional form of the ensemble remains the subject of recent debate \cite{blumenfeld-questions, tighe-gaussian, tighe-fluctuations, teitel-stress,sarkar}.

It is also unknown whether such statistical descriptions are appropriate, given that interparticle force distributions and probability densities in configuration space are known to be dependent on the loading protocol \cite{majmudar-photoelastic, bertrand}. Numerous systems, granular \cite{jamming-history-dependence, paulsen, jaeger-memory} and otherwise \cite{friction-history-dependence, sethna}, encode system history. 
Furthermore, shear-jamming \cite{bi-shear, vinutha} is distinct from compressive jamming \cite{jamming}, with the jamming transition occurring over a range of packing fractions and with additional states of minimum shear stress. 
While some properties of angoricity have been observed in shear-jammed systems \cite{bi-shear-angoricity}, there has been no systematic comparison across loading conditions.
 
To address these questions, we report measurements of particle positions and force-moment tensors for clusters of particles within experiments on a photoelastic granular material; this allows us to measure vector contact forces at the particle scale.
We subsequently extract the granular variables of interest in the force-moment ensemble by making measurements of angoricity for separate tensorial components and loading protocols. 
We find that this statistical mechanics framework must conserve both the force-moment tensor and the Maxwell-Cremona force-tiling area \cite{tighe-fluctuations}, leading to two protocol-dependent and one protocol-independent conjugate variables.

% THEORY
\paragraph*{Force-moment ensemble.--}
By analogy to the equilibrium thermodynamic energy, we calculate the global force-moment tensor \cite{statmech}, $\hat{\Sigma} = \sum_N {\vec r}_{\mu\nu}\otimes {\vec F}_{\mu\nu}$  over an experiment containing $N$ particles (enumerated $\mu,\nu$) with centers separated by a displacement ${\vec r}_{\mu\nu}$ and interparticle contact force ${\vec F}_{\mu\nu}$.
States of $\hat{\Sigma}$ are enumerated by the number of configurations $\Omega(\hat{\Sigma})$ and associated configurational entropy, $S=\ln{\Omega}$. 
The tensorial angoricity is correspondingly defined to be $\alpha_{ij} = \partial S/\partial \Sigma_{ij}$, and we expect a Boltzmann-like probability of observing a local cluster of $m$ nearby particles with force-moment tensor $\hat{\sigma}$ within a jammed particulate bath at angoricity $\hat{\alpha}$, given by: $\mathcal{P}(\hat{\sigma}|\hat{\alpha}) = \Omega(\hat{\sigma})e^{-\hat{\alpha}:\hat{\sigma}}/Z(\hat{\alpha})$.

As an extension to this theory, we incorporate the effect of the Maxwell-Cremona diagram, which is formed by mapping contact forces to a tiling of contacting polygons, as a representation of static force balance \cite{tighe-fluctuations}.
The total area of the tiling, $a$, is a conserved quantity, as rearrangements in force correspond to transfers of area between polygons \cite{tighe-gaussian, sarkar}.
(Note: although only frictionless packings sustain strictly convex force-tiles, we observe 85\% of our tiles to be convex even though they are frictional.) 
To incorporate this additional conservation law, we consider an extended Boltzmann-like term to account for the probability of observing a cluster of local force-moment $\hat{\sigma}$ and tiling area $a$:
\begin{equation}
\mathcal{P}(\hat{\sigma},a|\hat{\alpha},\kappa) = \frac{\Omega(\hat{\sigma},a)}{Z(\hat{\alpha},\kappa)} \exp{(-\hat{\alpha}:\hat{\sigma}-\kappa a)},
\label{distribution}
\end{equation}
where $\hat{\alpha}$ is the global angoricity associated with $\hat{\Sigma}$ and $Z(\hat{\alpha}, \kappa)$ is the partition function \cite{tighe-gaussian, teitel-stress}. 
We futher define $\kappa = \partial S/\partial a$ as the Lagrange multiplier associated with $a$, herein referred to as keramicity (from the Greek word  $\kappa \grave{\epsilon} \rho \alpha \mu o \varsigma$
meaning tile). %κέραμος from Daphne

\begin{figure}
\includegraphics[width=0.9\linewidth]{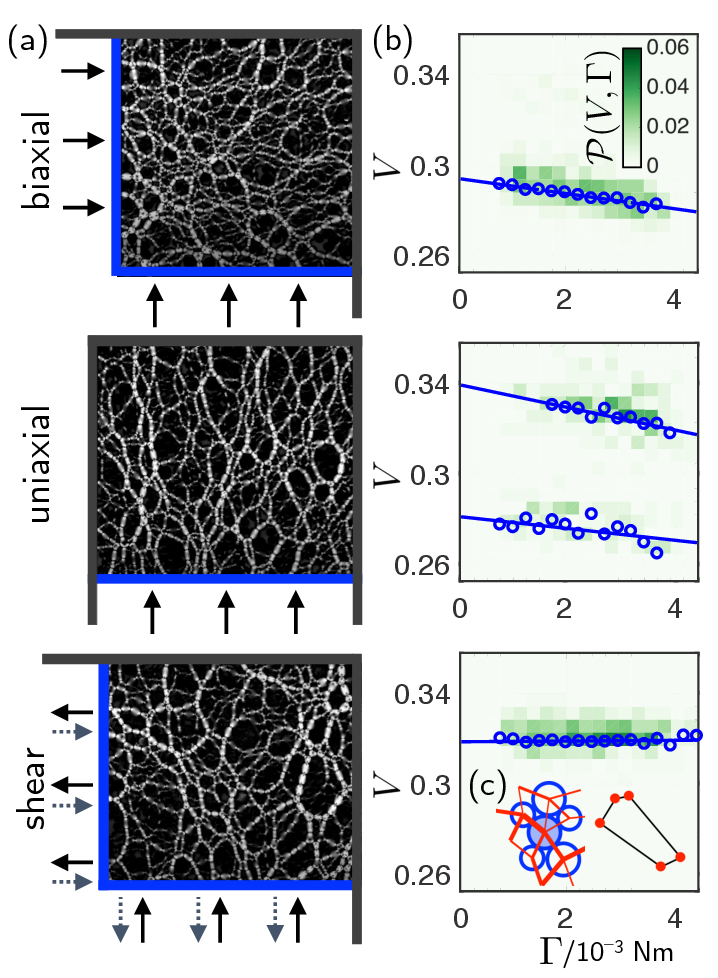}
\caption{
(a) Example photoelastic images at $\Gamma = 2.1\times 10^{-3}$~Nm for each of three loading schemes -- biaxial, uniaxial, and shear$^\mathrm{A}$-- shown schematically according to which walls move. Shear$^\mathrm{B}$ (dashed lines) is the opposite of shear$^\mathrm{A}$ (solid lines). Bright particles are under larger force; large-scale light gradients have been removed by image-processing.
(b) Joint probability distributions $\mathcal{P}(V,\Gamma)$ for dimensionless mean local free volume $V$ (averaged over clusters of size $m=8$) and global confining pressure $\Gamma$ for each protocol. Dashed lines are linear fits to $V(\Gamma)$ using a weighted average. (c) A force-balanced particle within a jammed packing has an associated Maxwell-Cremona tile, formed by a closed loop of all vector forces acting on a particle.
}
\label{f:protocols}
\end{figure}

% EXPERIMENT
\paragraph*{Experimental details.--}
We conduct experiments on a  two-dimensional  granular system composed of $N=890$ bidisperse  (Vishay PhotoStress PSM-4) circular discs with an equal fraction of radii $R_s = 5.5\,\mathrm{mm}$ and $R_l = 7.7\,\mathrm{mm}$ ($R_l/R_s=1.4$). A subfluidizing upflow of air passing through a porous polypropylene sheet floats the particles to eliminate basal friction. A rigid support grid ensures that the surface is level to reduce drift, and smooth to prevent particle clustering. Experimental details are provided in \cite{puckett}.

Each granular configuration is subject to quasistatic loading under one of three protocols: (1) uniaxial compression, (2) biaxial compression, and (3) simple shear, shown in Fig.~\ref{f:protocols}. 
Under all protocols, the discs are initially confined to a square region of $48\times 48\,\mathrm{cm}$ bounded by two walls controlled by stepper motors, and two fixed walls. Under uniaxial compression, a single motorized wall compresses the system with a series of steps of constant size ($0.6\,\mathrm{mm}$), while under biaxial compression, both motorized walls compress the system with the same step size. 
The loading walls then retract and the system dilates to a loose packing, activating a brief overheard flow of turbulent air which rearranges the particles to produce a new random configuration.
Under simple shear, after slowly compressing to approach the jamming transition, one wall moves in while another moves out. We conduct the shear experiments under the interchange of the two mobile walls to isolate the effect of any anisotropy on the surface (denoted by shear$^\mathrm{A}$ and shear$^\mathrm{B}$). 

We repeat the experiment to obtain at least 1,000 packings under each protocol, and for each packing capture images taken under polarized light, which contain the photoelastic response to interparticle forces. We track particle positions using MATLAB and use Voro++, a radical Vorono\"{i} tessellation tool, to determine disc neighbors and local volumes \cite{voro}. 
From the polarized images, we determine the normal and tangential contact forces using non-linear least-squares minimization between a numerically constructed fringe pattern and the actual image \cite{pegs, pegs-code}. 
This provides the forces $\vec{F}_{\mu\nu}$ and relative positions $\vec{r}_{\mu\nu}$ needed to construct the force-moment tensor. The full dataset will be available for download at \cite{data}.

% VOLUME-PRESSURE
 Local free volumes, preferred to global values for particle-scale statistical mechanics \cite{makse-phase, aste, mcnamara}, are calculated for clusters of $m=8$ particles, using the $m-1$ particles nearest to each central particle. To account for the particle bidispersity \cite{lechenault}, we determine the local free volume of a cluster from $v=\frac{1}{m} \sum_{i}[v_i-v_\mathrm{min}(R_i)]/v_\mathrm{min}(R_s)$
where $R_i$ is $R_s$ (or $R_l$) for small (or large) particles, $v_i$ is the corresponding particle Vorono\"{i} volume, and $v_\mathrm{min}=\sqrt{12}R_i^2$ is the smallest possible Vorono\"i volume.
Averaging the coarse-grained values of $v$ yields a global measure, $V$.
Figure~\ref{f:protocols}b demonstrates that histograms of the mean local free volume $V$, and global confining pressure $\Gamma = \frac{1}{N}\mathrm{Tr}(\hat{\Sigma})$ vary depending on the loading protocol. Note that this ``pressure'' has units of Nm instead of the usual 2D pressure N/m$^2$ since we have chosen, for simplicity, not to divide by the particle area.

We observe that across all three protocols there are two characteristic mean free volumes:
$V \approx 0.28$ observed for biaxial loading and 
$V \approx 0.32$ observed for shear loading. 
The looser-packed value for sheared samples is expected from Reynolds (or shear) dilatancy \cite{ren, granular-media-book, tighe-dilatancy}.
States prepared by uniaxial compression exhibit both values, in line with principal component analysis \cite{pca}.
Along each $V(\Gamma)$ characteristic curve, we observe $V$ decreases approximately linearly as a function of $\Gamma$, except in the case of shear, which by experimental design was constrained to constant $V$.

\paragraph*{Measuring temperaturelike quantities.--} 
In the absence of an explicitly known partition function, we measure the temperaturelike quantities $({\hat \alpha}, \kappa)$ using the method of overlapping histograms \cite{overlapping,dean-OH}, as has previously been applied to granular ensembles \cite{teitel-stress, henkes, puckett, mcnamara}.
We construct histograms of the components of $\hat{\sigma}$ over the entire range of  $(V,\Gamma)$ shown in Fig.~\ref{f:protocols}, binning states by the global measure $\Gamma$. We neglect data with $\Gamma < 10^{-3}\,\mathrm{Nm}$,  set by the lower limit of our force resolution.  From within each packing, we randomly generate clusters of size $m=8$ from nearby discs.
For each cluster, the force-moment tensor is computed additively over the particles in the cluster, and can be resolved into a normal $p = (\sigma_1+\sigma_2)/2$ and deviatoric $\tau = (\sigma_1-\sigma_2)/2$ components for eigenvalues $\sigma_1$ and $\sigma_2$.
As observed by \cite{henkes, puckett}, $m=8$ is large enough that measurements of temperature-like variables are independent of $m$.

\begin{figure}
\includegraphics[width=\linewidth]{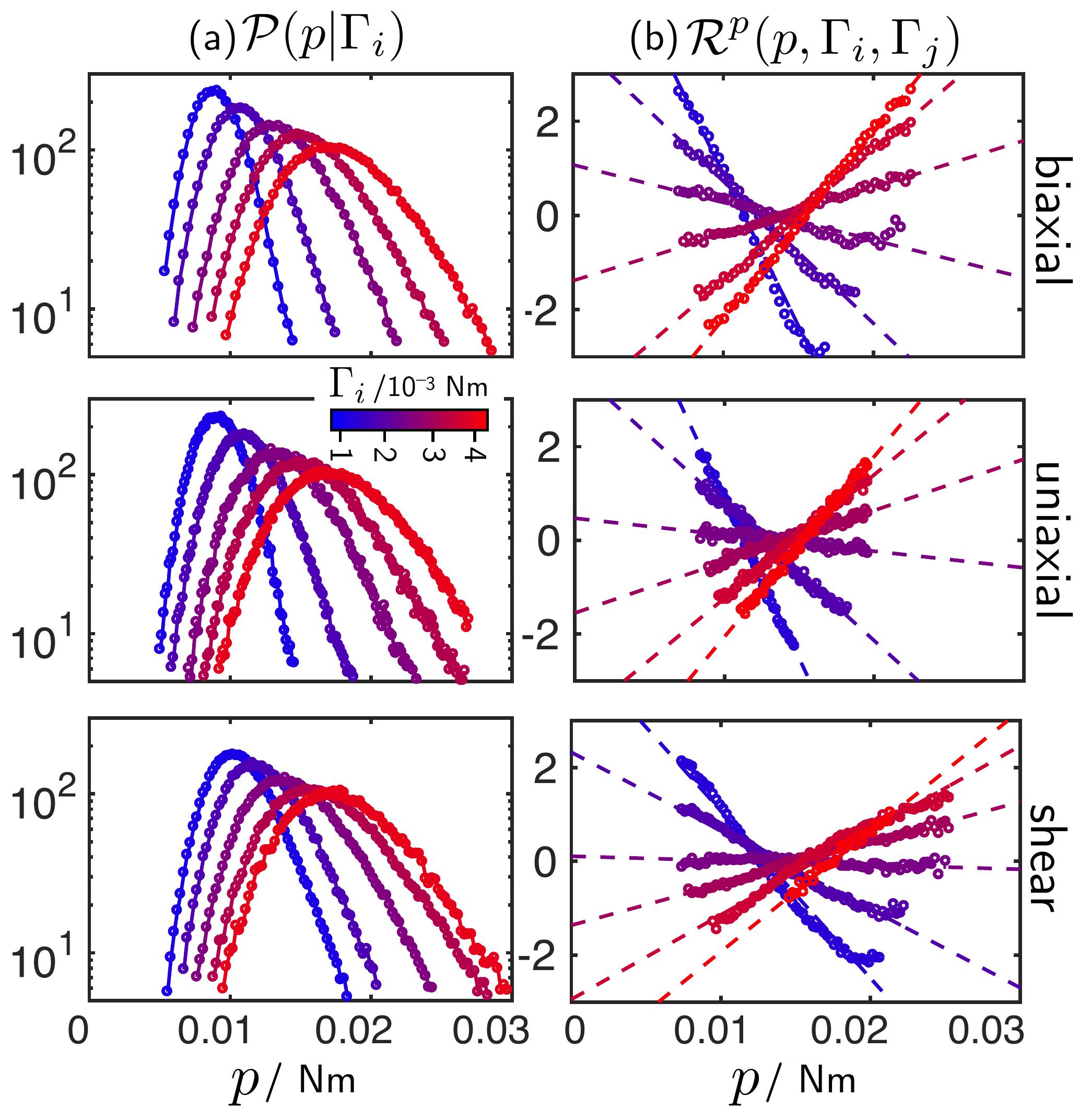}
\caption{(a) Histograms of local normal force-moment component $p$ on clusters of  $m=8$ particles, at fixed $\Gamma_i$, for all three loading protocols (shear$^\mathrm{A}$ and shear$^\mathrm{B}$ combined) (b) Associated histogram log-ratios $\mathcal{R}$ taken between probability distributions observed at confining pressure $\Gamma_i$ and $\Gamma_{j_0} = \frac{1}{2}\mathrm{max}(\Gamma_i)$. Each $\mathcal{R}(p)$ is  fit to a straight line with slope $\alpha^p_{j}-\alpha^p_{i}$ (Eq.~\ref{normalOH}, with $\kappa_j-\kappa_i \approx 0$).}
\label{f:Normal}
\end{figure}

\begin{figure}
\includegraphics[width=\linewidth]{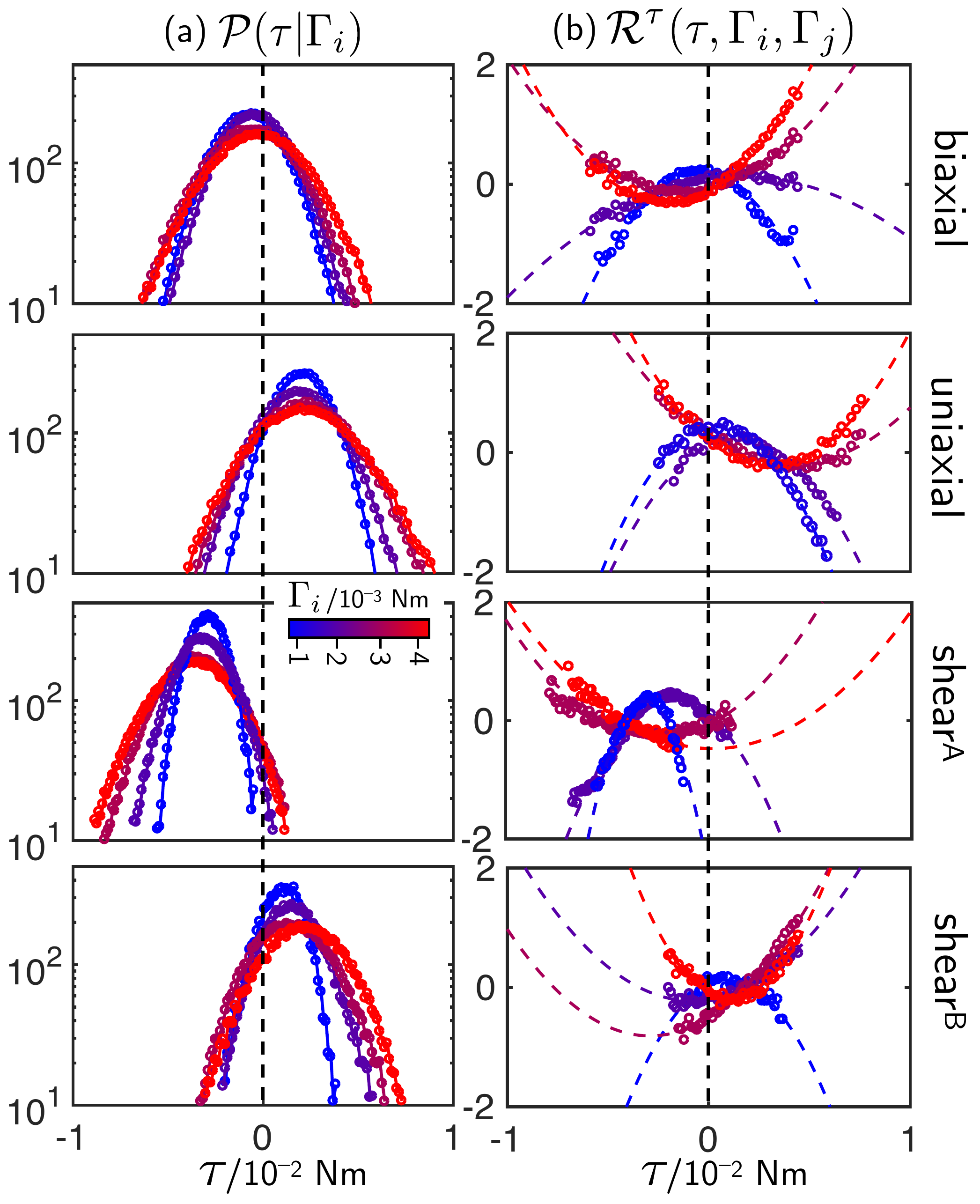}
\caption{(a) Histograms of local deviatoric force-moment component $\tau$ on clusters of  $m=8$ particles, at fixed $\Gamma_i$, for all four loading protocols (shear$^\mathrm{A}$ and shear$^\mathrm{B}$ separated) (b) Associated histogram log-ratios $\mathcal{R}$ taken between probability distributions observed at confining pressure $\Gamma_i$ and $\Gamma_{j_0} = \frac{1}{2}\mathrm{max}(\Gamma_i)$. Each $\mathcal{P}(\tau)$ is fit to a parabola with coefficients $\alpha^\tau_{j}-\alpha^\tau_{i}$ and $\kappa_j-\kappa_i$ (Eq.~\ref{devOH}).}
\label{f:Deviatoric}
\end{figure}

Under each of the four protocols, we compute the two associated distributions of the force-moment tensor, $\mathcal{P}(p|\Gamma)$ and $\mathcal{P}(\tau|\Gamma)$.
As shown in Fig.~\ref{f:Normal}-\ref{f:Deviatoric}, we observe that as confining pressure increases, the distributions of both normal and deviatoric components broaden \cite{teitel-stress, henkes, puckett, blumenfeld-questions}.
Note that the states prepared under the two different shear-loading protocols (A, B) were found to exhibit spontaneous symmetry breaking, resulting in a bias towards negative and positive deviatoric components, respectively \cite{nguyen}. Consequently, we treat these two protocols separately. For $\mathcal{P}(p)$ in both shear protocols, the mean and variance are indistinguishable from each other, while for $\mathcal{P}(\tau)$, the mean is translated in opposite directions.

% Figure: Deviatoric Stress

To proceed with the overlapping histogram methods, we 
utilize the ratio defined as $\mathcal{R}^\sigma(\sigma,\Gamma_i,\Gamma_j) = \log{\left[\mathcal{P}(\sigma|\Gamma_i)/\mathcal{P}(\sigma|\Gamma_j)\right]}$ for $\sigma \in \{p,\tau\}$. 
To give this ratio more meaning, 
consider the distribution of Eq.~\ref{distribution} under the assumption that the local force tiling area $a$ is strongly peaked at $\mathrm{det}(\hat{\sigma})$ \cite{teitel-stress,tighe-gaussian, tighe-fluctuations}. In this limit, we can integrate Eq.~\ref{distribution} to isolate distributions:
\begin{align}
\mathcal{P}(p|\Gamma) &= \frac{\tilde{\Omega}(p,\Gamma)}{Z(\hat{\alpha},\kappa)}\exp{(-\alpha^pp- \kappa p^2)},\label{normal} \\ 
\mathcal{P}(\tau|\Gamma) &=  \frac{\hat{\Omega}(\tau,\Gamma)}{Z(\hat{\alpha},\kappa)}\exp{(-\alpha^\tau\tau+ \kappa \tau^2)}, \label{deviatoric}
\end{align}
for modified densities of state $\bar{\Omega}$ and $\tilde{\Omega}$ (see Supplemental Material for detailed integration). Thus, we have three temperature-like quantities: $\alpha^p$ and $\alpha^\tau$, the normal and deviatoric components of the angoricity tensor, and $\kappa$, the keramicity. By taking the logarithm ratios of these two distributions (Eq.~\ref{normal}-\ref{deviatoric}) are
\begin{align}
\mathcal{R}^p(p,\Gamma_i,\Gamma_j) &= \mathcal{R}_0^p + (\alpha^p_{j}-\alpha^p_{i})p +  (\kappa_j-\kappa_i)p^2, \label{normalOH} \\
\mathcal{R}^\tau(\tau,\Gamma_i,\Gamma_j) &= \mathcal{R}_0^\tau + (\alpha^\tau_{j}-\alpha^\tau_{i})\tau - (\kappa_j-\kappa_i)\tau^2,
\label{devOH}
\end{align}
where $\mathcal{R}_0 = \log{(Z_i/Z_j)}$, and $(i,j)$ denotes two sets of states of pressure $\Gamma_i$ and $\Gamma_j$. Because ratios are sensitive to counting error in small values of the denominator, we only examine data for which $\mathcal{P} > 10$.

% Figure: Temperatures
\begin{figure*}[t]
\includegraphics[width=0.85\linewidth]{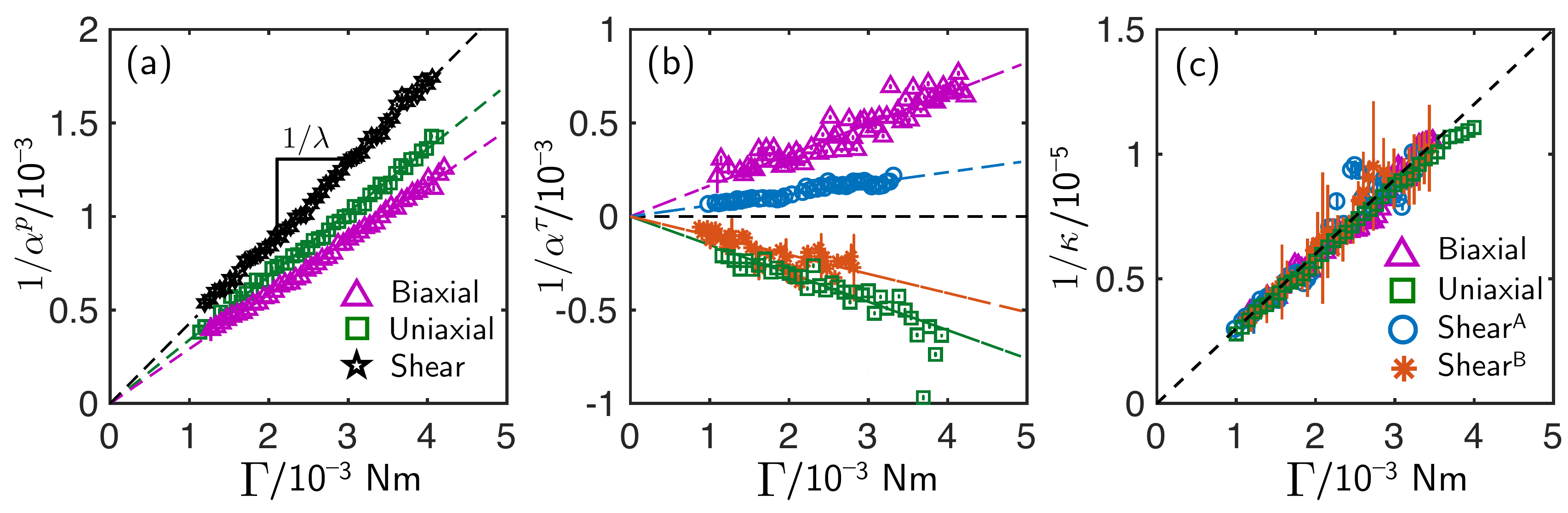}
\caption{Temperature-like variables: (a) Normal inverse angoricity, $1/\alpha^p$ against confining pressure $\Gamma$, as computed for three loading conditions [$\alpha^p\Gamma = 0.29\pm 0.01$ (biaxial), $0.34\pm 0.01$ (uniaxial), $0.43\pm 0.01$ (shear)]. (b) Deviatoric inverse angoricity, $1/\alpha^\tau$ plotted against the same, indicating opposite dependencies on confining pressure based on the loading protocol [$\alpha^\tau\Gamma = -1.5 \pm 0.1$ (uniaxial), $1.6\pm 0.1$ (biaxial), $0.59\pm0.03$ (shear$^\mathrm{A}$), $-1.0\pm0.3$ (shear$^\mathrm{B}$)]. (c) The inverse keramicity ($1/\kappa$) evolves linearly with confining pressure and does so with equal proportionality under all protocols [$\kappa\Gamma= (3.0\pm 0.1)\times 10^{-3}$]. Dashed lines represent linear fits.}
\label{f:Temperatures}
\end{figure*}

In Fig.~\ref{f:Normal}b and Fig.~\ref{f:Deviatoric}b, we illustrate typical logarithm ratios for several representative histograms; all ratios are computed relative to the same  reference histogram. 
We observe that $\mathcal{R}^p$ is linear in $p$ with some arbitrary offset associated with the ratio of partition functions. The slope of $\mathcal{R}^p(p)$  gives the differential normal angoricity, $\alpha^p_j-\alpha^p_i$ between states at confining pressures $\Gamma_j$ and $\Gamma_i$.
However, these ratios (Fig.~\ref{f:Normal}b) can only be fit to reasonable precision to a first order polynomial in $p$, such that the difference $\kappa_j-\kappa_i$ cannot be established from this approach.

In contrast, similar ratios computed in Fig.~\ref{f:Deviatoric}b are found to be parabolic with a non-zero second-order coefficient, matching the expectation from Eq.~\ref{devOH}, and we fit these ratios to a polynomial: $x\tau^2 + y\tau + z$ The fit parameter $x$, corresponding to changes in the variance of force-moment component distributions, gives $x=\kappa_i-\kappa_j$, the difference in keramicity between states of global confining pressure $\Gamma_i$ and $\Gamma_j$. As before, we also identify the fit parameter $y$ to be associated with $y=\alpha^\tau_j-\alpha^\tau_i$. 
Thus, by taking the logarithm of the ratio of two closely-overlapped histograms with sufficient proximity in local free volume \cite{henkes}, we can compute relative measures of $\alpha^p$, $\alpha^\tau$, and $\kappa$ to within an additive constant $\mathcal{R}_0$ \cite{puckett}.

The presence of a $\tau^2$ term is found for all three protocols, in agreement with predictions by \citet{teitel-stress}, in which the
multipliers $\alpha^p$ and $\kappa$ are strictly related to only global measures of pressure. This allows the Boltzmann-like factor found in Eq.~\ref{distribution} to include terms that are quadratic in the components of the force-moment tensor.

\paragraph*{Equations of state.--}  As shown for angoricity under isotropic compression, we propose that the computed temperature-like state variables can be described by equations of state of the form $\alpha \Gamma = \lambda$ \cite{field-theory, henkes, puckett}, where $\alpha$ represents any of $(\alpha^p, \alpha^\tau, \kappa)$ and $\lambda$ is a constant.
To test experimental adherence to this hypothesis, we identify $\frac{1}{\alpha}=0$ (and  $\frac{1}{\kappa}=0$) as corresponding to the jamming point ($\Gamma = 0$) and reference differential temperature-like variable measurements accordingly (see Supplemental Material for procedure).

Repeating this procedure  for all three temperature-like variables, we observe that the equation of state holds in all cases (Fig.~\ref{f:Temperatures}). 
While for $\kappa$ the constant $\lambda$ is independent of loading protocol, for both $\alpha^p$ and $\alpha^\tau$ the constant exhibits significant protocol-dependence. 
Interestingly, $\lambda$ is always positive for $\alpha^p$, but the sign varies for $\alpha^\tau$ (values given in caption), corresponding to the direction of translation of $\mathcal{P}(\tau)$ in Fig.~\ref{f:Deviatoric}a, some of which are associated with experimental asymmetries.
The magnitudes of the $\lambda$ agree within 2 standard errors for shear$^\mathrm{A}$ vs. shear$^\mathrm{B}$, and the signs differ due to the change in sign of $\tau$.
Because both components of $\hat{\alpha}$ are observed to be protocol-dependent, the tensoral angoricity is not valid as a variable of {\it state}, but rather a variable of {\it process.} By contrast, the keramicity behaves as a state variable. 

Angoricity measures a granular packing's ability to accommodate configurational rearrangements of equivalent global force-moment tensor (equivalently, stress) on a component-wise basis, $\alpha_{ij} = \partial S/\partial \Sigma_{ij}$. 
Thus, we have shown that as packings are loaded above jamming, they gain more and more configurations from the presence of forces to be distributed (Fig.~\ref{f:Temperatures}a). 
Meanwhile, as $\frac{1}{\alpha^p}$ (Fig.~\ref{f:Temperatures}a) and $\frac{1}{\alpha^\tau}$ (Fig.~\ref{f:Temperatures}b) grow in magnitude, the responsiveness of the number of configurations to increases in global stress ($\partial S/\partial \Sigma$) decreases.
Finally, the protocol by which a granular packing is loaded does not impact the trend that more compressed states gain fewer additional configurations from equivalent additions to the global Maxwell-Cremona area tiling (Fig.~\ref{f:Temperatures}c).

\paragraph*{Conclusion.--}
We have demonstrated non-negligible curvature in the ratio of overlapped force-moment tensor histograms in jammed states of granular systems. 
This observation points towards a formulation of the force-moment ensemble that requires conservation of Maxwell-Cremona area tiling. 
The associated temperature-like variables all grow inversely with respect to confining pressure, but the keramicity, associated with the force tiling, is the only valid variable of state. 
The path-dependence of variables describing soft matter systems is not unfamiliar, but protocol-dependent scaling could be a route to a comprehensive granular equation of state.
Meanwhile, the generalization of the force-moment ensemble to account for force-tiling statistics may yield a comprehensive statistical mechanics framework.

\paragraph*{Acknowledgements.--}
We are grateful for support from the National Science Foundation (DMR-1206808) and James S. McDonnell Foundation, and for useful conversations with Bob Behringer, Rafi Blumenfeld, Bulbul Chakraborty, Daphne Klotsa, James Puckett, Steve Teitel, and Brian Tighe.

% Bibliography
%\bibliographystyle{apsrev.bst}
%\bibliography{HistoryDependencePaper}

\begin{thebibliography}{39}
\expandafter\ifx\csname natexlab\endcsname\relax\def\natexlab#1{#1}\fi
\expandafter\ifx\csname bibnamefont\endcsname\relax
  \def\bibnamefont#1{#1}\fi
\expandafter\ifx\csname bibfnamefont\endcsname\relax
  \def\bibfnamefont#1{#1}\fi
\expandafter\ifx\csname citenamefont\endcsname\relax
  \def\citenamefont#1{#1}\fi
\expandafter\ifx\csname url\endcsname\relax
  \def\url#1{\texttt{#1}}\fi
\expandafter\ifx\csname urlprefix\endcsname\relax\def\urlprefix{URL }\fi
\providecommand{\bibinfo}[2]{#2}
\providecommand{\eprint}[2][]{\url{#2}}

\bibitem[{\citenamefont{Edwards and Oakeshott}(1989)}]{powders}
\bibinfo{author}{\bibfnamefont{S.~F.} \bibnamefont{Edwards}} \bibnamefont{and}
  \bibinfo{author}{\bibfnamefont{R.~B.~S.} \bibnamefont{Oakeshott}},
  \bibinfo{journal}{Physica A} \textbf{\bibinfo{volume}{157}},
  \bibinfo{pages}{1080} (\bibinfo{year}{1989}).

\bibitem[{\citenamefont{Jaeger and Nagel}(1992)}]{jaeger-overview}
\bibinfo{author}{\bibfnamefont{H.~M.} \bibnamefont{Jaeger}} \bibnamefont{and}
  \bibinfo{author}{\bibfnamefont{S.~R.} \bibnamefont{Nagel}},
  \bibinfo{journal}{Science} \textbf{\bibinfo{volume}{255}},
  \bibinfo{pages}{1523} (\bibinfo{year}{1992}).

\bibitem[{\citenamefont{Ball and Blumenfeld}(2002)}]{ball}
\bibinfo{author}{\bibfnamefont{R.~C.} \bibnamefont{Ball}} \bibnamefont{and}
  \bibinfo{author}{\bibfnamefont{R.}~\bibnamefont{Blumenfeld}},
  \bibinfo{journal}{Physical Review Letters} \textbf{\bibinfo{volume}{88}},
  \bibinfo{pages}{115505} (\bibinfo{year}{2002}).

\bibitem[{\citenamefont{Martiniani et~al.}(2017)\citenamefont{Martiniani,
  Schrenk, Ramola, Chakraborty, and Frenkel}}]{martiniani}
\bibinfo{author}{\bibfnamefont{S.}~\bibnamefont{Martiniani}},
  \bibinfo{author}{\bibfnamefont{K.~J.} \bibnamefont{Schrenk}},
  \bibinfo{author}{\bibfnamefont{K.}~\bibnamefont{Ramola}},
  \bibinfo{author}{\bibfnamefont{B.}~\bibnamefont{Chakraborty}},
  \bibnamefont{and} \bibinfo{author}{\bibfnamefont{D.}~\bibnamefont{Frenkel}},
  \bibinfo{journal}{Nature Physics}  (\bibinfo{year}{2017}).

\bibitem[{\citenamefont{Bi et~al.}(2015)\citenamefont{Bi, Henkes, Daniels, and
  Chakraborty}}]{statmech}
\bibinfo{author}{\bibfnamefont{D.}~\bibnamefont{Bi}},
  \bibinfo{author}{\bibfnamefont{S.}~\bibnamefont{Henkes}},
  \bibinfo{author}{\bibfnamefont{K.~E.} \bibnamefont{Daniels}},
  \bibnamefont{and}
  \bibinfo{author}{\bibfnamefont{B.}~\bibnamefont{Chakraborty}},
  \bibinfo{journal}{Annual Review of Condensed Matter Physics}
  \textbf{\bibinfo{volume}{6}}, \bibinfo{pages}{63} (\bibinfo{year}{2015}).

\bibitem[{\citenamefont{Henkes et~al.}(2007)\citenamefont{Henkes, O'Hern, and
  Chakraborty}}]{henkes}
\bibinfo{author}{\bibfnamefont{S.}~\bibnamefont{Henkes}},
  \bibinfo{author}{\bibfnamefont{C.~S.} \bibnamefont{O'Hern}},
  \bibnamefont{and}
  \bibinfo{author}{\bibfnamefont{B.}~\bibnamefont{Chakraborty}},
  \bibinfo{journal}{Physical Review Letters} \textbf{\bibinfo{volume}{99}},
  \bibinfo{pages}{038002} (\bibinfo{year}{2007}).

\bibitem[{\citenamefont{Puckett and Daniels}(2013)}]{puckett}
\bibinfo{author}{\bibfnamefont{J.~G.} \bibnamefont{Puckett}} \bibnamefont{and}
  \bibinfo{author}{\bibfnamefont{K.~E.} \bibnamefont{Daniels}},
  \bibinfo{journal}{Physical Review Letters} \textbf{\bibinfo{volume}{110}},
  \bibinfo{pages}{058001} (\bibinfo{year}{2013}).

\bibitem[{\citenamefont{Blumenfeld and Edwards}(2009)}]{blumenfeld-questions}
\bibinfo{author}{\bibfnamefont{R.}~\bibnamefont{Blumenfeld}} \bibnamefont{and}
  \bibinfo{author}{\bibfnamefont{S.~F.} \bibnamefont{Edwards}},
  \bibinfo{journal}{Journal of Physical Chemistry B}
  \textbf{\bibinfo{volume}{113}}, \bibinfo{pages}{3981} (\bibinfo{year}{2009}).

\bibitem[{\citenamefont{Tighe et~al.}(2008)\citenamefont{Tighe, van Eerd, and
  Vlugt}}]{tighe-gaussian}
\bibinfo{author}{\bibfnamefont{B.~P.} \bibnamefont{Tighe}},
  \bibinfo{author}{\bibfnamefont{A.~R.~T.} \bibnamefont{van Eerd}},
  \bibnamefont{and} \bibinfo{author}{\bibfnamefont{T.~J.~H.}
  \bibnamefont{Vlugt}}, \bibinfo{journal}{Physical Review Letters}
  \textbf{\bibinfo{volume}{100}}, \bibinfo{pages}{238001}
  (\bibinfo{year}{2008}).

\bibitem[{\citenamefont{Tighe and Vlugt}(2011)}]{tighe-fluctuations}
\bibinfo{author}{\bibfnamefont{B.~P.} \bibnamefont{Tighe}} \bibnamefont{and}
  \bibinfo{author}{\bibfnamefont{T.~J.~H.} \bibnamefont{Vlugt}},
  \bibinfo{journal}{Journal of Statistical Mechanics: Theory and Experiment}
  \textbf{\bibinfo{volume}{4}}, \bibinfo{pages}{04002} (\bibinfo{year}{2011}).

\bibitem[{\citenamefont{Wu and Teitel}(2015)}]{teitel-stress}
\bibinfo{author}{\bibfnamefont{Y.}~\bibnamefont{Wu}} \bibnamefont{and}
  \bibinfo{author}{\bibfnamefont{S.}~\bibnamefont{Teitel}},
  \bibinfo{journal}{Physical Review E} \textbf{\bibinfo{volume}{92}},
  \bibinfo{pages}{022207} (\bibinfo{year}{2015}).

\bibitem[{\citenamefont{Sarkar et~al.}(2016)\citenamefont{Sarkar, Bi, Zhang,
  Ren, Behringer, and Chakraborty}}]{sarkar}
\bibinfo{author}{\bibfnamefont{S.}~\bibnamefont{Sarkar}},
  \bibinfo{author}{\bibfnamefont{D.}~\bibnamefont{Bi}},
  \bibinfo{author}{\bibfnamefont{J.}~\bibnamefont{Zhang}},
  \bibinfo{author}{\bibfnamefont{J.}~\bibnamefont{Ren}},
  \bibinfo{author}{\bibfnamefont{R.~P.} \bibnamefont{Behringer}},
  \bibnamefont{and}
  \bibinfo{author}{\bibfnamefont{B.}~\bibnamefont{Chakraborty}},
  \bibinfo{journal}{Physical Review E} \textbf{\bibinfo{volume}{93}},
  \bibinfo{pages}{042901} (\bibinfo{year}{2016}).

\bibitem[{\citenamefont{Majmudar and Behringer}(2005)}]{majmudar-photoelastic}
\bibinfo{author}{\bibfnamefont{T.~S.} \bibnamefont{Majmudar}} \bibnamefont{and}
  \bibinfo{author}{\bibfnamefont{R.~P.} \bibnamefont{Behringer}},
  \bibinfo{journal}{Nature} \textbf{\bibinfo{volume}{435}}
  (\bibinfo{year}{2005}).

\bibitem[{\citenamefont{Bertrand et~al.}(2016)\citenamefont{Bertrand,
  Behringer, Chakraborty, O'Hern, and Shattuck}}]{bertrand}
\bibinfo{author}{\bibfnamefont{T.}~\bibnamefont{Bertrand}},
  \bibinfo{author}{\bibfnamefont{R.~P.} \bibnamefont{Behringer}},
  \bibinfo{author}{\bibfnamefont{B.}~\bibnamefont{Chakraborty}},
  \bibinfo{author}{\bibfnamefont{C.~S.} \bibnamefont{O'Hern}},
  \bibnamefont{and} \bibinfo{author}{\bibfnamefont{M.~D.}
  \bibnamefont{Shattuck}}, \bibinfo{journal}{Physical Review E}
  \textbf{\bibinfo{volume}{93}}, \bibinfo{pages}{012901}
  (\bibinfo{year}{2016}).

\bibitem[{\citenamefont{Chaudhuri et~al.}(2010)\citenamefont{Chaudhuri,
  Berthier, and Sastry}}]{jamming-history-dependence}
\bibinfo{author}{\bibfnamefont{P.}~\bibnamefont{Chaudhuri}},
  \bibinfo{author}{\bibfnamefont{L.}~\bibnamefont{Berthier}}, \bibnamefont{and}
  \bibinfo{author}{\bibfnamefont{S.}~\bibnamefont{Sastry}},
  \bibinfo{journal}{Physical Review Letters} \textbf{\bibinfo{volume}{104}},
  \bibinfo{pages}{165701} (\bibinfo{year}{2010}).

\bibitem[{\citenamefont{Paulsen et~al.}(2014)\citenamefont{Paulsen, Keim, and
  Nagel}}]{paulsen}
\bibinfo{author}{\bibfnamefont{J.~D.} \bibnamefont{Paulsen}},
  \bibinfo{author}{\bibfnamefont{N.~C.} \bibnamefont{Keim}}, \bibnamefont{and}
  \bibinfo{author}{\bibfnamefont{S.~R.} \bibnamefont{Nagel}},
  \bibinfo{journal}{Physical Review Letters} \textbf{\bibinfo{volume}{113}},
  \bibinfo{pages}{068301} (\bibinfo{year}{2014}).

\bibitem[{\citenamefont{Josserand et~al.}(2000)\citenamefont{Josserand,
  Tkachenko, Mueth, and Jaeger}}]{jaeger-memory}
\bibinfo{author}{\bibfnamefont{C.}~\bibnamefont{Josserand}},
  \bibinfo{author}{\bibfnamefont{A.~V.} \bibnamefont{Tkachenko}},
  \bibinfo{author}{\bibfnamefont{D.~M.} \bibnamefont{Mueth}}, \bibnamefont{and}
  \bibinfo{author}{\bibfnamefont{H.~M.} \bibnamefont{Jaeger}},
  \bibinfo{journal}{Physical Review Letters} \textbf{\bibinfo{volume}{85}},
  \bibinfo{pages}{3632} (\bibinfo{year}{2000}).

\bibitem[{\citenamefont{Rubinstein et~al.}(2006)\citenamefont{Rubinstein,
  Cohen, and Fineberg}}]{friction-history-dependence}
\bibinfo{author}{\bibfnamefont{S.~M.} \bibnamefont{Rubinstein}},
  \bibinfo{author}{\bibfnamefont{G.}~\bibnamefont{Cohen}}, \bibnamefont{and}
  \bibinfo{author}{\bibfnamefont{J.}~\bibnamefont{Fineberg}},
  \bibinfo{journal}{Physical Review Letters} \textbf{\bibinfo{volume}{96}},
  \bibinfo{pages}{256103} (\bibinfo{year}{2006}).

\bibitem[{\citenamefont{Huang and Sethna}(1991)}]{sethna}
\bibinfo{author}{\bibfnamefont{M.}~\bibnamefont{Huang}} \bibnamefont{and}
  \bibinfo{author}{\bibfnamefont{J.~P.} \bibnamefont{Sethna}},
  \bibinfo{journal}{Physical Review B} \textbf{\bibinfo{volume}{43}},
  \bibinfo{pages}{3245} (\bibinfo{year}{1991}).

\bibitem[{\citenamefont{Bi et~al.}(2011)\citenamefont{Bi, Zhang, Chakraborty,
  and Behringer}}]{bi-shear}
\bibinfo{author}{\bibfnamefont{D.}~\bibnamefont{Bi}},
  \bibinfo{author}{\bibfnamefont{J.}~\bibnamefont{Zhang}},
  \bibinfo{author}{\bibfnamefont{B.}~\bibnamefont{Chakraborty}},
  \bibnamefont{and} \bibinfo{author}{\bibfnamefont{R.~P.}
  \bibnamefont{Behringer}}, \bibinfo{journal}{Nature}
  \textbf{\bibinfo{volume}{480}}, \bibinfo{pages}{335} (\bibinfo{year}{2011}).

\bibitem[{\citenamefont{Vinutha and Sastry}(2016)}]{vinutha}
\bibinfo{author}{\bibfnamefont{H.~A.} \bibnamefont{Vinutha}} \bibnamefont{and}
  \bibinfo{author}{\bibfnamefont{S.}~\bibnamefont{Sastry}},
  \bibinfo{journal}{Nature Physics} \textbf{\bibinfo{volume}{12}},
  \bibinfo{pages}{578} (\bibinfo{year}{2016}).

\bibitem[{\citenamefont{Liu and Nagel}(2010)}]{jamming}
\bibinfo{author}{\bibfnamefont{A.~J.} \bibnamefont{Liu}} \bibnamefont{and}
  \bibinfo{author}{\bibfnamefont{S.~R.} \bibnamefont{Nagel}},
  \bibinfo{journal}{Annual Review of Condensed Matter Physics}
  \textbf{\bibinfo{volume}{1}} (\bibinfo{year}{2010}).

\bibitem[{\citenamefont{Bi et~al.}(2013)\citenamefont{Bi, Zhang, Behringer, and
  Chakraborty}}]{bi-shear-angoricity}
\bibinfo{author}{\bibfnamefont{D.}~\bibnamefont{Bi}},
  \bibinfo{author}{\bibfnamefont{J.}~\bibnamefont{Zhang}},
  \bibinfo{author}{\bibfnamefont{R.~P.} \bibnamefont{Behringer}},
  \bibnamefont{and}
  \bibinfo{author}{\bibfnamefont{B.}~\bibnamefont{Chakraborty}},
  \bibinfo{journal}{Europhysics Letters} \textbf{\bibinfo{volume}{102}},
  \bibinfo{pages}{34002} (\bibinfo{year}{2013}).

\bibitem[{\citenamefont{Rycroft et~al.}(2006)\citenamefont{Rycroft, Grest,
  Landry, and Bazant}}]{voro}
\bibinfo{author}{\bibfnamefont{C.~H.} \bibnamefont{Rycroft}},
  \bibinfo{author}{\bibfnamefont{G.~S.} \bibnamefont{Grest}},
  \bibinfo{author}{\bibfnamefont{J.~W.} \bibnamefont{Landry}},
  \bibnamefont{and} \bibinfo{author}{\bibfnamefont{M.~Z.}
  \bibnamefont{Bazant}}, \bibinfo{journal}{Physical Review E}
  \textbf{\bibinfo{volume}{74}}, \bibinfo{pages}{021306}
  (\bibinfo{year}{2006}).

\bibitem[{\citenamefont{Daniels et~al.}(2017)\citenamefont{Daniels, Kollmer,
  and Puckett}}]{pegs}
\bibinfo{author}{\bibfnamefont{K.~E.} \bibnamefont{Daniels}},
  \bibinfo{author}{\bibfnamefont{J.~E.} \bibnamefont{Kollmer}},
  \bibnamefont{and} \bibinfo{author}{\bibfnamefont{J.~G.}
  \bibnamefont{Puckett}}, \bibinfo{journal}{Review of Scientific Instruments}
  \textbf{\bibinfo{volume}{88}}, \bibinfo{pages}{051808}
  (\bibinfo{year}{2017}).

\bibitem[{\citenamefont{Kollmer}()}]{pegs-code}
\bibinfo{author}{\bibfnamefont{J.~E.} \bibnamefont{Kollmer}},
  \emph{\bibinfo{title}{Photo-Elastic Grain Solver}},
  \bibinfo{address}{{\url{http://github.com/jekollmer/PEGS}}}.

\bibitem[{dat()}]{data}
\emph{\bibinfo{title}{Link to data at \url{http://DataDryad.org} will go
  here.}}

\bibitem[{\citenamefont{Song et~al.}(2008)\citenamefont{Song, Wang, and
  Makse}}]{makse-phase}
\bibinfo{author}{\bibfnamefont{C.}~\bibnamefont{Song}},
  \bibinfo{author}{\bibfnamefont{P.}~\bibnamefont{Wang}}, \bibnamefont{and}
  \bibinfo{author}{\bibfnamefont{H.~A.} \bibnamefont{Makse}},
  \bibinfo{journal}{Nature} \textbf{\bibinfo{volume}{453}},
  \bibinfo{pages}{629} (\bibinfo{year}{2008}).

\bibitem[{\citenamefont{Aste and {Di Matteo}}(2008)}]{aste}
\bibinfo{author}{\bibfnamefont{T.}~\bibnamefont{Aste}} \bibnamefont{and}
  \bibinfo{author}{\bibfnamefont{T.}~\bibnamefont{{Di Matteo}}},
  \bibinfo{journal}{Physical Review E} \textbf{\bibinfo{volume}{77}},
  \bibinfo{pages}{021309} (\bibinfo{year}{2008}).

\bibitem[{\citenamefont{McNamara et~al.}(2009)\citenamefont{McNamara, Richard,
  de~Richter, Ca{\"e}r, and Delannay}}]{mcnamara}
\bibinfo{author}{\bibfnamefont{S.}~\bibnamefont{McNamara}},
  \bibinfo{author}{\bibfnamefont{P.}~\bibnamefont{Richard}},
  \bibinfo{author}{\bibfnamefont{S.~K.} \bibnamefont{de~Richter}},
  \bibinfo{author}{\bibfnamefont{G.} \bibnamefont{Le Ca{\"e}r}},
  \bibnamefont{and} \bibinfo{author}{\bibfnamefont{R.}~\bibnamefont{Delannay}},
  \bibinfo{journal}{Physical Review E} \textbf{\bibinfo{volume}{80}},
  \bibinfo{pages}{031301} (\bibinfo{year}{2009}).

\bibitem[{\citenamefont{Lechenault et~al.}(2006)\citenamefont{Lechenault,
  da~Cruz, Dauchot, and Bertin}}]{lechenault}
\bibinfo{author}{\bibfnamefont{F.}~\bibnamefont{Lechenault}},
  \bibinfo{author}{\bibfnamefont{F.}~\bibnamefont{da~Cruz}},
  \bibinfo{author}{\bibfnamefont{O.}~\bibnamefont{Dauchot}}, \bibnamefont{and}
  \bibinfo{author}{\bibfnamefont{E.}~\bibnamefont{Bertin}},
  \bibinfo{journal}{Journal of Statistical Mechanics: Theory and Experiment}
  \textbf{\bibinfo{volume}{2006}}, \bibinfo{pages}{07009}
  (\bibinfo{year}{2006}).

\bibitem[{\citenamefont{Ren et~al.}(2013)\citenamefont{Ren, Dijksman, and
  Behringer}}]{ren}
\bibinfo{author}{\bibfnamefont{J.}~\bibnamefont{Ren}},
  \bibinfo{author}{\bibfnamefont{J.~A.} \bibnamefont{Dijksman}},
  \bibnamefont{and} \bibinfo{author}{\bibfnamefont{R.~P.}
  \bibnamefont{Behringer}}, \bibinfo{journal}{Physical Review Letters}
  \textbf{\bibinfo{volume}{110}}, \bibinfo{pages}{018302}
  (\bibinfo{year}{2013}).

\bibitem[{\citenamefont{Andreotti et~al.}(2013)\citenamefont{Andreotti,
  Forterre, and Pouliquen}}]{granular-media-book}
\bibinfo{author}{\bibfnamefont{B.}~\bibnamefont{Andreotti}},
  \bibinfo{author}{\bibfnamefont{Y.}~\bibnamefont{Forterre}}, \bibnamefont{and}
  \bibinfo{author}{\bibfnamefont{O.}~\bibnamefont{Pouliquen}},
  \emph{\bibinfo{title}{Granular media: between fluid and solid}}
  (\bibinfo{publisher}{Cambridge University Press}, \bibinfo{year}{2013}).

\bibitem[{\citenamefont{Tighe}(2014)}]{tighe-dilatancy}
\bibinfo{author}{\bibfnamefont{B.~P.} \bibnamefont{Tighe}},
  \bibinfo{journal}{Granular Matter} \textbf{\bibinfo{volume}{16}}
  (\bibinfo{year}{2014}).

\bibitem[{\citenamefont{Hotelling}(1933)}]{pca}
\bibinfo{author}{\bibfnamefont{H.}~\bibnamefont{Hotelling}},
  \emph{\bibinfo{title}{Analysis of a Complex of Statistical Variables into
  Principal Components}} (\bibinfo{publisher}{Warwick \& York},
  \bibinfo{year}{1933}).

\bibitem[{\citenamefont{Labastie and Whetten}(1990)}]{overlapping}
\bibinfo{author}{\bibfnamefont{P.}~\bibnamefont{Labastie}} \bibnamefont{and}
  \bibinfo{author}{\bibfnamefont{R.~L.} \bibnamefont{Whetten}},
  \bibinfo{journal}{Physical Review Letters} \textbf{\bibinfo{volume}{65}},
  \bibinfo{pages}{1567} (\bibinfo{year}{1990}).

\bibitem[{\citenamefont{Dean and Lef\`{e}vre}(2003)}]{dean-OH}
\bibinfo{author}{\bibfnamefont{D.~S.} \bibnamefont{Dean}} \bibnamefont{and}
  \bibinfo{author}{\bibfnamefont{A.}~\bibnamefont{Lef\`{e}vre}},
  \bibinfo{journal}{Physical Review Letters} \textbf{\bibinfo{volume}{90}},
  \bibinfo{pages}{198301} (\bibinfo{year}{2003}).

\bibitem[{\citenamefont{Nguyen and Amon}(2016)}]{nguyen}
\bibinfo{author}{\bibfnamefont{T.~B.} \bibnamefont{Nguyen}} \bibnamefont{and}
  \bibinfo{author}{\bibfnamefont{A.}~\bibnamefont{Amon}},
  \bibinfo{journal}{Europhysics Letters} \textbf{\bibinfo{volume}{116}},
  \bibinfo{pages}{28007} (\bibinfo{year}{2016}).

\bibitem[{\citenamefont{Henkes and Chakraborty}(2005)}]{field-theory}
\bibinfo{author}{\bibfnamefont{S.}~\bibnamefont{Henkes}} \bibnamefont{and}
  \bibinfo{author}{\bibfnamefont{B.}~\bibnamefont{Chakraborty}},
  \bibinfo{journal}{Physical Review Letters} \textbf{\bibinfo{volume}{95}},
  \bibinfo{pages}{198002} (\bibinfo{year}{2005}).

\end{thebibliography}

\newpage
\onecolumngrid
\appendix

\centerline{Supplemental Material}
\centerline{Protocol-Dependence and State Variables in the Force-Moment Ensemble} 
\centerline{Ephraim S. Bililign, Jonathan E. Kollmer, Karen E. Daniels}
\centerline{Department of Physics, North Carolina State University, Raleigh, North Carolina 27695, USA}

\section{Integrating the Generalized Boltzmann-like Distribution}

In the extended force-moment ensemble, there are two conserved quantities for isotropically jammed states: $\hat{\Sigma}$ and $A$, the Maxwell-Cremona force-tiling area. Thus, the microcanonical ensemble probabilities are generalized to incorporate cluster force-moment tensor $\hat{\sigma}$ and cluster force-tiling area, $a$ in the following manner:
\begin{equation*}
\mathcal{P}(\hat{\sigma},a|\hat{\alpha},\kappa) = \frac{\Omega(\hat{\sigma},a)}{Z(\hat{\alpha},\kappa)}\exp{(-\hat{\alpha}: \hat{\sigma} - \kappa a)},
\end{equation*}
where $\kappa$ is some additional intensive variable of state necessary in this ensemble, which we have named {\it keramicity}. This assumes entropy is maximized with respect to the two Lagrange multipliers $\hat{\alpha}$ and $\kappa$, the former being reducible to the invariants: $\alpha^p$ and $\alpha^\tau$, which follows from the generalized result
\begin{equation*}
\hat{\alpha}:\hat{\sigma} = \mathrm{Tr}(\hat{\alpha}\otimes \hat{\sigma}) = p(\alpha_{xx}+\alpha_{yy})+\tau(\alpha_{xx}-\alpha_{yy})+2\alpha_{xy}\sigma_{xy} \approx p\alpha^p + \tau\alpha^\tau,\\
\end{equation*}
where we neglect the relatively small contribution of off-diagonal terms, $\sigma_{xy}$, in the local force-moment tensor and define $\alpha^p\equiv \alpha_{xx}+\alpha_{yy}$ and $\alpha^\tau\equiv \alpha_{xx}-\alpha_{yy}$. For instance, for all biaxial states, we find $\braket{\sigma_{xy}^2/\sigma_{xx}\sigma_{yy}} \approx 0.05$. As proposed by Wu and Teitel in Ref.~\cite{teitel-stress} and Tighe in Ref.~\cite{tighe-gaussian}, this leads to a general form of the probability distribution for observing a cluster of force-moment tensor $\hat{\sigma}$ and force-tiling area $a$ within a system of global ``temperature" $\hat{\alpha}$ and $\kappa$. Note that these variables are said to be functions of $\Gamma$, an invariant of the isotropic force-moment tensor, $\Gamma = \frac{1}{N}\mathrm{Tr}\hat{\Sigma}$. We begin with the result
\begin{equation}
\mathcal{P}(\hat{\sigma},a|\hat{\alpha},\kappa) = \frac{\Omega(\hat{\sigma},a)}{Z(\hat{\alpha},\kappa)}\exp{(-\alpha^pp-\alpha^\tau\tau - \kappa a)}.
\label{general-result}
\end{equation}

To get distributions of only $p$ or $\tau$, two easily measurable quantities, we assume that the fluctuations in  $a$ are strongly peaked at $\braket{a}\propto \mathrm{det}(\hat{\sigma}) = p^2-\tau^2$, then integrating Eq.~\ref{general-result} over $a$ gives
\begin{align*}
\mathcal{P}(\hat{\sigma}|\hat{\alpha},\kappa) &= \int da\, \mathcal{P}(\hat{\sigma},a|\hat{\alpha},\kappa) \\
&= \int da\,\frac{\Omega(\hat{\sigma},a)}{Z(\hat{\alpha},\kappa)}\exp{(-\alpha^pp-\alpha^\tau\tau - \kappa a)}\\
& = \frac{\bar{\Omega}(\hat{\sigma})}{Z(\hat{\alpha},\kappa)}\exp{[-\alpha^pp-\alpha^\tau\tau - \kappa(p^2-\tau^2)]},
\end{align*}
where $\bar{\Omega}$ is an effective density of states related to the counting of microstates with force-moment $\hat{\sigma}$:
\begin{equation*}
\bar{\Omega}(\hat{\sigma}) = \int da\, \Omega(\hat{\sigma},a).
\end{equation*}
We can now integrate with respect to $\sigma_{xy}$ to reduce $\bar{\Omega}\to \tilde{\Omega}(p,\tau)$, a function of the normal and deviatoric force-moment component on a cluster. To reduce the distribution one step further, we have two options:

\begin{enumerate}
\item Integrating with respect to $\tau$ after integrating away $\sigma_{xy}$ gives us $\mathcal{P}(p|\Gamma)$. That is,
\begin{align} \nonumber
\mathcal{P}(p|\hat{\alpha},\kappa) &= \int d\tau \int d\sigma_{xy}\mathcal{P}(\hat{\sigma}|\hat{\alpha},\kappa)\\ \nonumber
&= \int d\tau \int d\sigma_{xy}\frac{\bar{\Omega}(\hat{\sigma})}{Z(\hat{\alpha},\kappa)}\exp{[-\alpha^pp-\alpha^\tau\tau - \kappa(p^2-\tau^2)]}\\
&= \frac{\tilde{\Omega}(p)}{\bar{Z}(\hat{\alpha},\kappa)}\exp{(-\alpha^pp-\kappa p^2)},
\end{align}
where the reduced density of states absorbs dependence on $\tau$ and $\sigma_{xy}$,
\begin{equation*}
\tilde{\Omega}(p) = \iint d\tau\,d\sigma_{xy} \bar{\Omega}(\hat{\sigma})\exp{(-\alpha^\tau\tau+\kappa\tau^2)},
\end{equation*}
and is therefore a function of both $\alpha^\tau$ and $\kappa$, and accordingly a functional of $\Gamma$. 
Then, the method of overlapping histograms, when performed for distributions of normal force-moment component measured at two different confining pressures $\Gamma_i$ and $\Gamma_j$ gives
\begin{equation}
\mathcal{R}^p(p,\Gamma_i,\Gamma_j) = \log{\left[\frac{\mathcal{P}(p|\Gamma_i)}{\mathcal{P}(p|\Gamma_j)}\right]}= \mathcal{R}_0 + (\alpha^p_j-\alpha^p_i)p + (\kappa_j-\kappa_i)p^2,
\end{equation}
where $\mathcal{R}_0$ is some additive factor that accounts for the logarithm ratio of the partition functions, $\log{(Z_i/Z_j)}$. If $\kappa \ll \alpha$, then this would lead to the expected form of the logarithm ratio reported above.

\item Alternatively, integrating with respect to $p$ after integrating away $\sigma_{xy}$ gives us $\mathcal{P}(\tau|\Gamma)$. That is,
\begin{align} \nonumber
\mathcal{P}(\tau|\hat{\alpha},\kappa) &= \int dp \int d\sigma_{xy}\mathcal{P}(\hat{\sigma}|\hat{\alpha},\kappa)\\ \nonumber
&= \int dp \int d\sigma_{xy}\frac{\bar{\Omega}(\hat{\sigma})}{Z(\hat{\alpha},\kappa)}\exp{[-\alpha^pp-\alpha^\tau\tau - \kappa(p^2-\tau^2)]}\\ 
&= \frac{\hat{\Omega}(\tau)}{\bar{Z}(\hat{\alpha},\kappa)}\exp{(-\alpha^\tau\tau+\kappa \tau^2)},
\end{align}
where the reduced density of states absorbs dependence on $p$ and $\sigma_{xy}$,
\begin{equation*}
\hat{\Omega}(p) = \iint dp\,d\sigma_{xy} \bar{\Omega}(\hat{\sigma})\exp{(-\alpha^p p-\kappa p^2)},
\end{equation*}
and is therefore a function of both $\alpha^p$ and $\kappa$, and accordingly a functional of $\Gamma$.
The equivalent formulation of the method of overlapping histograms gives
\begin{equation}
\mathcal{R}^\tau(\tau,\Gamma_i,\Gamma_j) = \log{\left[\frac{\mathcal{P}(\tau|\Gamma_i)}{\mathcal{P}(\tau|\Gamma_j)}\right]} = \mathcal{R}_0 + (\alpha^\tau_j-\alpha^\tau_i)\tau + (\kappa_i-\kappa_j)\tau^2.
\end{equation}
Note that this is contingent on the assumption that $\hat{\Omega}$ is weakly dependent on $\Gamma$ so that $\hat\Omega(\tau,\Gamma_i)/\hat\Omega(\tau,\Gamma_j) \approx 1$, which is difficult to demonstrate in experiment.

\end{enumerate}

In summary, we are, using the method of overlapping histograms, in search of relationships for: $\alpha^p\Gamma$, $\alpha^\tau\Gamma$, and $\kappa\Gamma$.

\section{From Relative to Absolute Angoricity}
To compute absolute angoricity from a relative measure, we identify $\frac{1}{\alpha}=0$ as corresponding to the jamming transition. Therefore, $\alpha\to \infty$ as a granular packing becomes increasingly mechanically unconstrained. We define a matrix of differential angoricites $\alpha_{ij} \equiv \alpha_i-\alpha_j$, using the method of overlapping histograms. From this, we subtract an offset $\alpha_{0j}$, taken for $\Gamma_0$ corresponding to the smallest bin in confining pressure: $\alpha'_{ij} = \alpha_{ij}-\alpha_{0j}$. 
Averaging over all pressures, $\alpha'_i = \braket{\alpha'_{ij}}_j$. We then compute the parameter $\alpha_0$ for which the intercept in the linear least-squares fit of $1/(\alpha_i'-\alpha_0)$ to $\Gamma$ is minimized, giving us an absolute angoricity $\alpha(\Gamma_i) \equiv \alpha_i=\alpha_i'-\alpha_0$.

\section{Data Publication}

All raw datafiles (particle positions, interparticle vector forces) will be uploaded to \url{http://DataDryad.org} at the time of publication.

\end{document}